\documentclass[letter,prd,aps,twocolumn,floatfix,superscriptaddress,nofootinbib]{revtex4}
\pdfoutput=1

\usepackage{amssymb,amsmath,graphicx}
\usepackage{hyperref}
\usepackage[usenames,dvipsnames]{color}
\usepackage{array}
\makeatletter
\renewcommand*{\p@section}{\S\,}
\renewcommand*{\p@subsection}{\S\,}

\makeatother

\def\jnl@style{\it}
\def\aaref@jnl#1{{\jnl@style#1}}

\def\aaref@jnl#1{{\jnl@style#1}}

\def\aj{\aaref@jnl{AJ}}                   
\def\araa{\aaref@jnl{ARA\&A}}             
\def\apj{\aaref@jnl{ApJ}}                 
\def\apjl{\aaref@jnl{ApJ}}                
\def\apjs{\aaref@jnl{ApJS}}               
\def\ao{\aaref@jnl{Appl.~Opt.}}           
\def\apss{\aaref@jnl{Ap\&SS}}             
\def\aap{\aaref@jnl{A\&A}}                
\def\aapr{\aaref@jnl{A\&A~Rev.}}          
\def\aaps{\aaref@jnl{A\&AS}}              
\def\azh{\aaref@jnl{AZh}}                 
\def\baas{\aaref@jnl{BAAS}}               
\def\jrasc{\aaref@jnl{JRASC}}             
\def\memras{\aaref@jnl{MmRAS}}            
\def\mnras{\aaref@jnl{MNRAS}}             
\def\pra{\aaref@jnl{Phys.~Rev.~A}}        
\def\prb{\aaref@jnl{Phys.~Rev.~B}}        
\def\prc{\aaref@jnl{Phys.~Rev.~C}}        
\def\prd{\aaref@jnl{Phys.~Rev.~D}}        
\def\pre{\aaref@jnl{Phys.~Rev.~E}}        
\def\prl{\aaref@jnl{Phys.~Rev.~Lett.}}    
\def\pasp{\aaref@jnl{PASP}}               
\def\pasj{\aaref@jnl{PASJ}}               
\def\qjras{\aaref@jnl{QJRAS}}             
\def\skytel{\aaref@jnl{S\&T}}             
\def\solphys{\aaref@jnl{Sol.~Phys.}}      
\def\sovast{\aaref@jnl{Soviet~Ast.}}      
\def\ssr{\aaref@jnl{Space~Sci.~Rev.}}     
\def\zap{\aaref@jnl{ZAp}}                 
\def\nat{\aaref@jnl{Nature}}              
\def\iaucirc{\aaref@jnl{IAU~Circ.}}       
\def\aplett{\aaref@jnl{Astrophys.~Lett.}} 
\def\apspr{\aaref@jnl{Astrophys.~Space~Phys.~Res.}}
\def\bain{\aaref@jnl{Bull.~Astron.~Inst.~Netherlands}} 
\def\fcp{\aaref@jnl{Fund.~Cosmic~Phys.}}  
\def\gca{\aaref@jnl{Geochim.~Cosmochim.~Acta}}   
\def\grl{\aaref@jnl{Geophys.~Res.~Lett.}} 
\def\jcp{\aaref@jnl{J.~Chem.~Phys.}}      
\def\jgr{\aaref@jnl{J.~Geophys.~Res.}}    
\def\jqsrt{\aaref@jnl{J.~Quant.~Spec.~Radiat.~Transf.}}
\def\memsai{\aaref@jnl{Mem.~Soc.~Astron.~Italiana}}
\def\nphysa{\aaref@jnl{Nucl.~Phys.~A}}   
\def\physrep{\aaref@jnl{Phys.~Rep.}}   
\def\physscr{\aaref@jnl{Phys.~Scr}}   
\def\planss{\aaref@jnl{Planet.~Space~Sci.}}   
\def\procspie{\aaref@jnl{Proc.~SPIE}}   

\begin{document}

\date{\today}
\title{Probing Near-Horizon Fluctuations with Black Hole Binary Mergers}

\author{Steven L. Liebling}
\affiliation{Department of Physics, Long Island University, Brookville, New York 11548, USA}
\author{Matthew Lippert}
\author{Michael Kavic}
\affiliation{Department of Physics, Long Island University, Brooklyn, New York 11201, USA}


\begin{abstract}
The strong version of the nonviolent nonlocality proposal of Giddings predicts ``strong but soft" quantum metric fluctuations near black hole horizons in an attempt to resolve the information paradox. 
To study observable signatures of this proposal, we numerically solve Einstein's equations modified by these fluctuations and analyze the gravitational wave signal from the inspiral and merger of two black holes.
In a model of evolution for such fluctuations, we show that they lead to significant deviations in the observed waveform, even when the black holes are still well separated, and could potentially be observed by aLIGO.

\end{abstract}

\maketitle


\section{Introduction}\label{introduction}
The groundbreaking observations of binary black hole collisions by Advanced LIGO~(aLIGO) have ushered in a new era of gravitational wave astronomy~\cite{Abbott:2016blz,Abbott:2016nmj,Abbott:2017vtc}.  
The gravitational wave signal emitted from the inspiral, coalescence, and ringdown of two black holes permits direct observation of extreme physics in the near-horizon region.  Even the few
observations to date contain a wealth of information which can be used to test proposed modifications to the conventional general relativistic picture of a black hole~\cite{TheLIGOScientific:2016src}.

The semi-classical description of black hole collapse and evaporation leads to the information paradox, whose resolution has motivated proposals for radically new physics (see~\cite{Unruh:2017uaw} and \cite{Marolf:2017jkr} for two recent reviews).  If information is not irrevocably lost but somehow emerges from behind the horizon, the traditional description in terms of local quantum field theory on a semi-classical geometry must be modified, and not just by small, subtle effects~\cite{Page:1993wv, Almheiri:2012rt, Mathur:2005zp}.  A number of such drastic resolutions have been suggested; for example, the locally harmless horizon could be replaced with a hard and deadly firewall~\cite{Almheiri:2012rt} or the entire black hole geometry might be replaced by a non-geometric fuzzball~\cite{Mathur:2005zp}.

The nonviolent nonlocality proposal~\cite{Giddings:2011ks, Giddings:2012gc, Giddings:2014nla, Giddings:2014ova, Giddings:2016tla, Giddings:2017mym} (also \cite{Yosifov:2017pxj}), by contrast, does not eliminate the event horizon. 
Local low-energy quantum field theory outside the horizon is supplemented by nonlocal couplings with degrees of freedom inside the black hole, leading to soft quantum metric fluctuations whose wavelengths are not set by the Planck scale but instead by the much larger Schwarzschild radius.  
According the strong version of the proposal, these are unit-size fluctuations in order
to allow information to escape at the necessary rate.

These types of significant modifications to the standard black hole picture might be expected to have observable consequences. 
Much of the recent focus has been on the prospects for imaging the supermassive black hole Sagittarius~A* using the Event Horizon Telescope~\cite{Giddings:2014ova, Giddings:2016btb}. High-precision timing measurements from a suitable pulsar-black hole binary could also detect large-amplitude quantum fluctuations~\cite{Estes:2016wgv}.

However, the gravitational wave signal from a black hole binary merger is currently the most promising venue in which to hunt for signs of new physics and to constrain deviations from general relativity (GR). Several recent attempts have been made to model the gravitational wave predictions of various alternative theories (e.g.~\cite{emd, Okounkova:2017yby}).  One approach~\cite{Barcelo:2014cla, Cardoso:2016oxy, Maggio:2017ivp, 2017arXiv170606155M} is to replace the black hole with an object nearly as compact, a so-called ``exotic compact object" and explore any observational differences due solely to the removal of the horizon. In~\cite{Abedi:2016hgu}, the horizon was replaced with a reflecting boundary as a proxy for a firewall and faint echoes of the original signal were reported in the late stages of the ringdown.\footnote{These claims have been controversial; see~\cite{Ashton:2016xff, Abedi:2017isz, Barcelo:2017lnx, Price:2017cjr}.}

It should be noted that the requirement that the quantum fluctuations have order-one amplitudes has recently been questioned~\cite{Giddings:2017mym}.  In this weaker version of the proposal, the fluctuations of the geometry may be exponentially suppressed, rendering any astrophysical signals completely unobservable.  

In this paper, we focus on the original, stronger nonviolent nonlocality proposal and analyze the resulting modifications to the binary black hole merger signal.  We numerically evolve the Einstein equations for two equal-mass black holes, colliding them both head-on and in a quasi-circular inspiral, implementing the quantum effects as 
Schwarzschild-scale metric fluctuations, supported within roughly a Schwarzschild radius of the horizon. We vary the both the amplitude and frequency of the metric fluctuations.

In anticipating the observational effect of the nonviolent nonlocality scenario, we should emphasize (as noted by \cite{Giddings:2016tla}) that the proposed quantum fluctuations occur near the horizon, and so the metric well outside the black hole  might be expected to resemble that of a standard black hole without such fluctuations.  Similarly, the spacetime  outside a neutron star is the same as that of a black hole of equal mass.  Of course, the neutron star lacks the black hole's horizon, but this difference is relatively unimportant in a well separated binary system.  The distinction between black hole and neutron star binaries only becomes apparent when the constituents are sufficiently close that the structure of the objects begins to affect the dynamics.\footnote{See Damour's ``effacement theorem''~\cite{Hawking:1987en}.}  Because the modifications in the nonviolent nonlocality scenario are confined to the near-horizon region, this analogy with a neutron start suggests that the nonviolent nonlocality proposal would also have minimal dynamical implications until just before the merger.  However, the numerical results we present below suggest otherwise.

We find that the
fluctuations yield significant deviations from the baseline GR signal.  Within the frequency band where aLIGO is sensitive, we calculate the overlap between the merger waveform derived from GR and the one arising when the metric fluctuations are present. Our results imply that the strong version of the
nonviolent nonlocality proposal produces potentially observable effects for nominal choices of parameters.  However, deviations due to the fluctuations could conceivably be mimicked by changes in other parameters, such as black hole masses or spins.  Extensive further numerical studies would be needed to definitively rule out this possibility.

Contrary to the expectations of~\cite{Giddings:2016tla}, which argued that the effects of quantum fluctuations would appear only in the very late inspiral and merger process, we find these effects are already present during the early inspiral phase.  This suggests that the impact of the fluctuations radiates beyond their near-horizon region of support in a manner very different from a firewall or other exotic compact object.

\section{Numerical System}\label{system}
%
We begin with an existing code which makes use of
the BSSN formulation~\cite{1995PhRvD..52.5428S, 1999PhRvD..59b4007B} to implement the Einstein equations. 
This code has been tested without the fluctuations described here
extensively and used to study a number of different
phenomena~\cite{Palenzuela:2009yr,Neilsen:2010ax,Liebling:2016orx}.
In this approach, the spacetime metric is written as 
\begin{equation}
ds^2 = -\alpha^2\,dt^2 
+ \gamma_{ij}\left(dx^i + \beta^i\,dt\right)\left(dx^j + \beta^j\,dt\right),
\end{equation}
with $\alpha$ the lapse function and $\beta^i$ the shift vector.
Specifically, we express  the Einstein equations in terms of the BSSN-NOK
formalism~\cite{1987PThPS..90....1N,1995PhRvD..52.5428S,1999PhRvD..59b4007B,
lousto}.
In this formulation, the metric on spatial hypersurfaces
$\gamma_{ij}$ is expressed in terms of a conformal factor $\chi$
and a conformally flat metric $\tilde \gamma_{ij}$ as
\begin{equation}
\gamma_{ij} = \frac{1}{\chi}\tilde\gamma_{ij}, \qquad 
         \chi = \left(\det \gamma_{ij}\right)^{-1/3}
\end{equation}
such that $\det\tilde\gamma_{ij}=1$.

The initial data  consists of the superposition of two equal-mass black holes (for details, see \cite{Hannam:2009ib}).\footnote{Note that these black holes will quickly evolve consistent with the puncture method to have a ``trumpet'' topology and thus differ from Schwarzschild black holes; for details, see \cite{Hannam:2009ib}.}  We consider two types of initial momenta:  the black holes boosted directly toward each other leading to a head-on collision, and the black holes given opposite transverse boosts to achieve an approximately quasi-circular orbit. 
We work in units of the total mass of the system $M$. Each black hole, therefore, has mass roughly $M/2$.

We then modify the code to incorporate the metric fluctuations.  The nonviolent nonlocality scenario proposes a nonlocal coupling between operators inside the black hole and the exterior stress tensor \cite{Giddings:2012gc} which, when integrated out, acts as a quantum source for the metric.  In our code, this effect is represented as a fluctuating classical source perturbing the metric.  Because the source is supported in the near-horizon region, the location of the source must constantly be updated to track the moving black holes.

We take a simple, somewhat ad hoc implementation which captures the essential physics.  
At each step in the numerical integration, we modify by hand the time derivatives 
of the diagonal components of the metric\footnote{The $\pm H$ helps avoid signature changes of the metric as explained in Section~III of Ref.~\cite{Giddings:2016btb}.}
\begin{eqnarray}
      {\dot {\alpha}} & \rightarrow &  {\dot {\alpha }} \left(1+ H\right) \nonumber \\
      {\dot {\tilde \gamma}}_{ii} & \rightarrow &  {\dot {\tilde \gamma}}_{ii} \left(1- H\right)\ \ .
      \label{fluffimplementation}
 \end{eqnarray}
For the perturbation $H$, we postulate the following form
\begin{eqnarray}
H = A \exp^{-\left(\vec{r}-\vec{r}_{1|2}\right)^2/R^2}\sin\left(\Omega t\right) \ ,
\end{eqnarray}
where $A$ is the amplitude and $\Omega$ is the angular frequency.  We assume a Gaussian radial profile of width $R$ centered on the locations of the two black holes. The coordinate location of the nearer black hole is denoted by $\vec{r}_{1|2}$, and we track these locations by monitoring the minimum values of $\chi$.

The metric perturbation $H$ in the nonviolent nonlocality proposal has an order-one amplitude, has support within a region roughly a Schwarzschild radius from the horizon, and has a frequency on the order of the inverse light-crossing time of the black hole.\footnote{We work in units where Newton's constant $G = 1$.   For a black hole of mass $M/2$, the Schwarzschild radius is $M$.}   As the exact values of these parameters are not specified by the nonviolent nonlocality model, we will scan over a range such that:  $A \lesssim 1$, $R/M \approx 1$, and $M \Omega  \approx 1$.

We adopt a Courant parameter of $\lambda = 0.25$ using
a self-shadow adaptive mesh method with seven levels of dynamic refinement
achieving a finest resolution of $0.04 M$.

To extract physical information, we monitor the
gravitational Newman-Penrose radiative scalar
$\Psi_4$ on spherical shells located in the wave zone (typically with radius $80M$).  
This quantity is computed by contracting the Weyl tensor with a suitably defined null tetrad,
\begin{eqnarray}
  \Psi_4 = C_{abcd} n^a \bar m^b n^c \bar m^d  \ ,
\end{eqnarray}
and accounts for the energy carried off by outgoing waves at infinity.

The strain $h(t)$, which is directly observed by aLIGO, is related to the gravitational scalar by
\begin{eqnarray}
\frac{d^2}{dt^2} h = \Psi_4.
\end{eqnarray}


\section{Results}\label{results}
%
Despite the ad hoc nature of the modifications, the evolution is well behaved. 
We compute the $L^2$-norms of the Hamiltonian constraint and the momentum constraints. The residuals of the modified and GR evolutions are comparable, suggesting that our numerical scheme to modify gravity remains reasonable throughout the evolution.

\begin{figure}[h]
\centering
\includegraphics[width=9.0cm,angle=0]{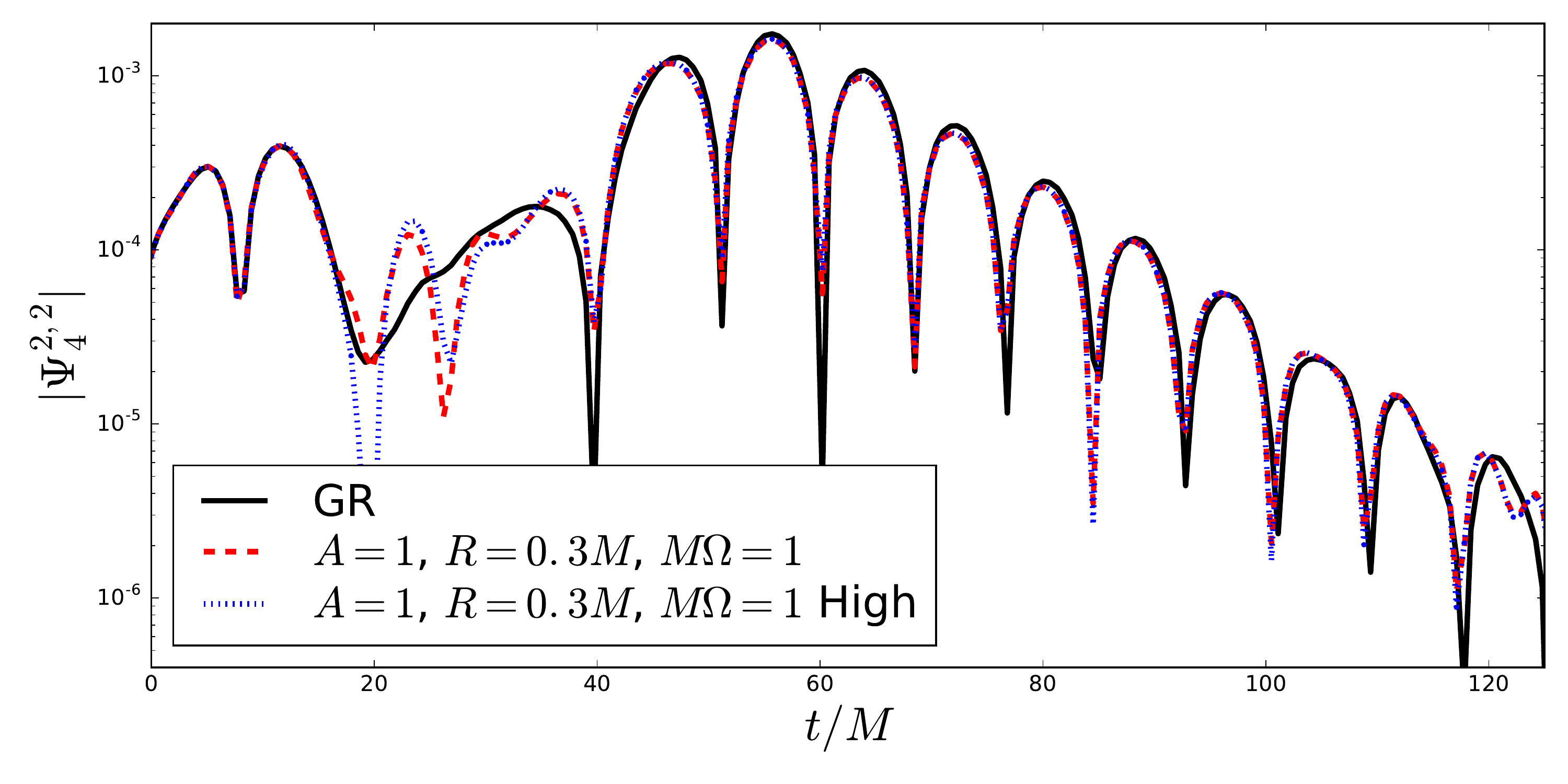}
\caption{ \textit{Signals from a head-on collision  for two resolutions.} Although not fully
converged because there is no continuum solution, the high- and low-resolution results
are very similar, suggesting the scheme is reasonably well-behaved. 
}
\label{fig:convergence}
\end{figure}

Second, we display in Fig.~\ref{fig:convergence} the gravitational wave signal due to a head-on collision with quantum fluctuations, computed at two different resolutions, along with the unmodified GR computation. It seems unlikely, although not impossible, the that added fluctuations \eqref{fluffimplementation} can originate from a discretized set of continuum differential equations. And, if we are not solving a system of differential equations with a unique, continuum solution, true convergence is not expected.  

However, if the results were highly dependent on resolution, it would be difficult to make predictions.  As indicated in the figure the results are, to a good approximation, independent of resolution.

Head-on collisions provide only a very short time window for quantum fluctuations to have an effect.  However, they do provide a measure of the quasi-normal modes of an out-of-equilibrium black hole, namely the remnant black hole as it settles to a quiescent state. As shown in Fig.~\ref{fig:convergence}, compared with the GR computation, the quantum fluctuations do result in differences in the brief, pre-merger signal, but the ringdown is largely the same for all the evolutions.

Note that the quantum fluctuations couple only to the dynamics, in that they are introduced as multiplicative
factors to the time derivatives of the metric \eqref{fluffimplementation}. Hence, after the merger, when the remnant is settling down, the time derivatives become small and thus so do these
modifications. One could alternatively consider some modification that coupled directly to the metric instead of its time derivatives.

The quasi-circular inspiral is more sensitive to quantum fluctuations than the head-on collision both because
it takes longer for the merger to occur and also because the breaking of axial symmetry allows more complex motion.
It was previously observed with the Einstein-Maxwell-dilaton system studied in~\cite{emd} that orbital
mergers were much more sensitive probes of deviations from GR than head-on collisions, and we find this to be true here as well.

In particular, quantum fluctuations in the neighborhood around the black holes
introduce eccentricity in the black hole orbits and produce
high frequency features in the gravitational wave signatures.

\begin{figure}[ht]
\centering
\includegraphics[width=9.0cm,angle=0]{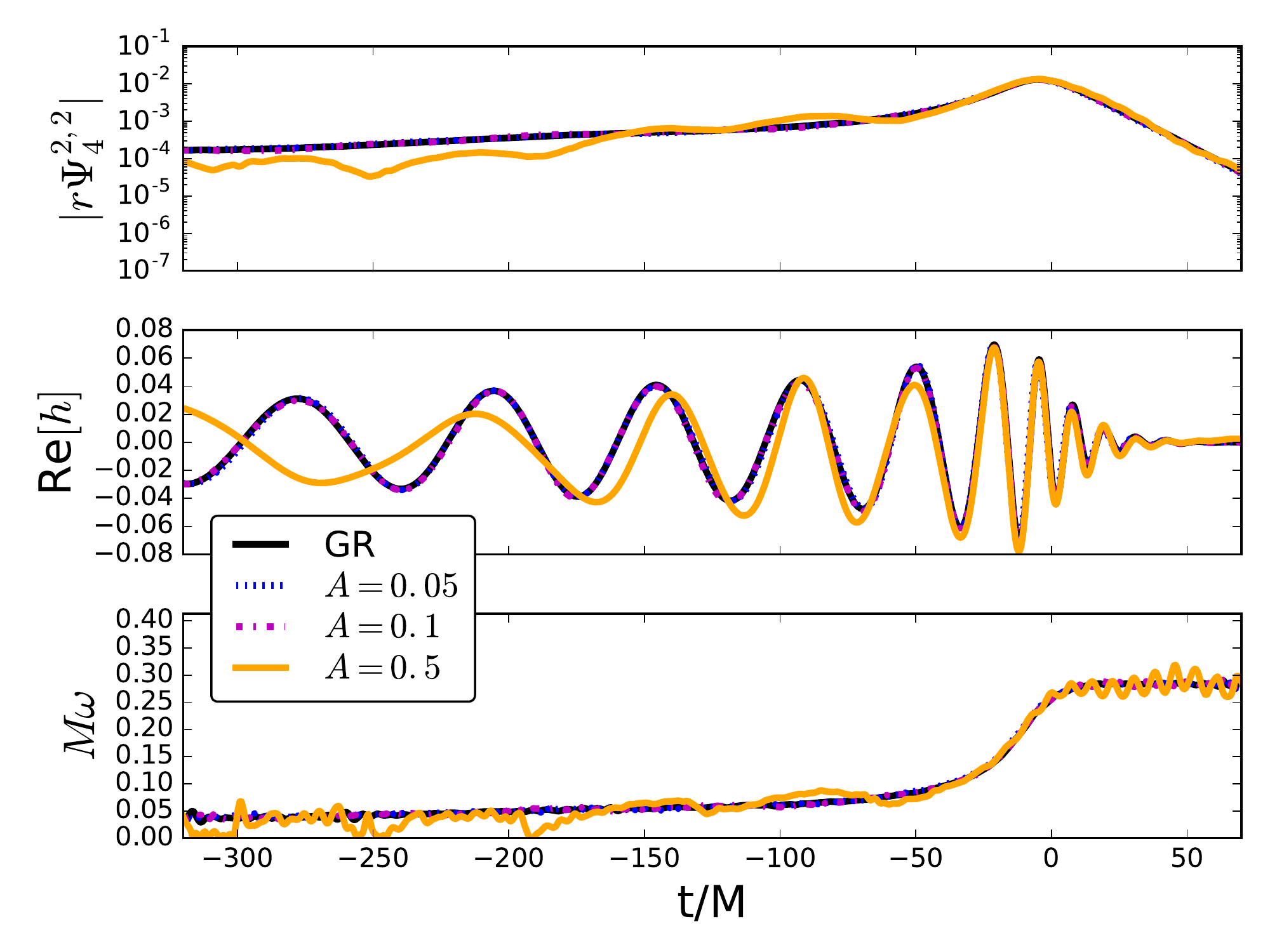}
\caption{ \textit{Results for the inspiral and coalescence of two, equal-mass black holes, for varying amplitudes $A$ of the quantum fluctuations.}
These runs have $M\Omega=0.1$ and $R_0=0.3M$. 
The times and overall phases of the nonviolent nonlocality waveforms have been shifted to
maximize the overlap with the GR waveform.
\textbf{Top:} The magnitude of the gravitational wave scalar $|r \Psi^{2,2}_4|$.
\textbf{Middle:} The real component of the strain $h$. 
\textbf{Bottom:} The angular frequency $\omega$ of the binary.
}
\label{fig:orbital}
\end{figure}

\begin{figure}[ht]
\centering
\includegraphics[width=9.0cm,angle=0]{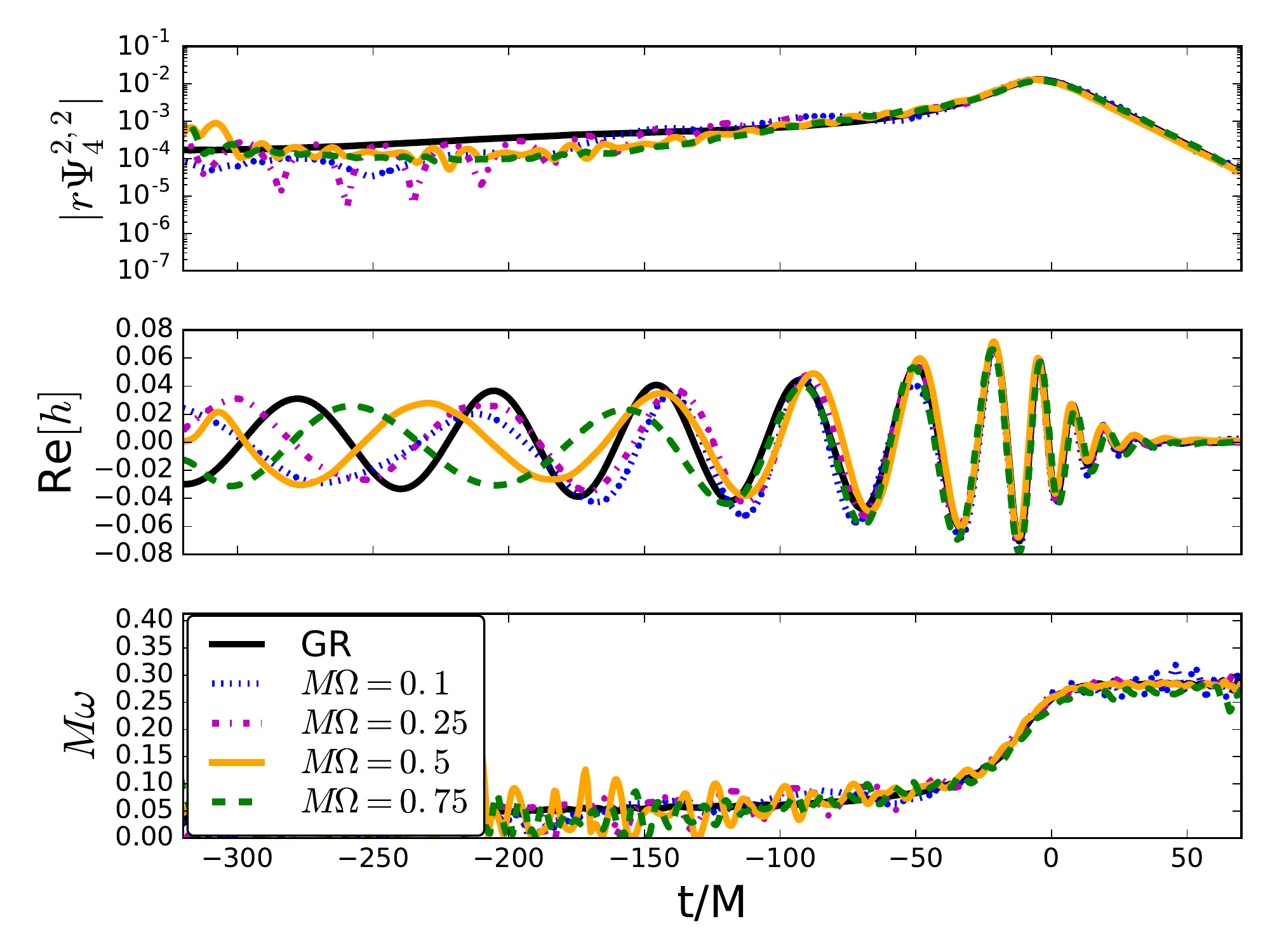}
\caption{ \textit{Results of varying the fluctuation frequency $\Omega$ for the inspiral and coalescence of two, equal-mass black holes.}
These runs have $A=0.5M$ and $R_0=0.3M$.
The same quantities as shown in Fig.~\ref{fig:orbital} are presented here,
and again the times and overall phases of the nonviolent nonlocality waveforms have been shifted to
maximize the overlap with the GR waveform.
}
\label{fig:orbitalvaryomega}
\end{figure}

We present results from orbital coalescences in Figs.~\ref{fig:orbital}
and~\ref{fig:orbitalvaryomega}, showing a range of fluctuation amplitudes $A$
and frequencies $\Omega$, respectively.
The three panels in each figure display the (dominant) $l=2=m$ component of the
radiative scalar $\Psi^{2,2}_4$ in the top panel, the real component of its corresponding wave strain $h$
in the middle panel, and the angular frequency $\omega$ of the binary as a function of time.
These figures illustrate significant differences between the nonviolent nonlocality model and the GR prediction.

\begin{figure}[ht]
\centering
\includegraphics[width=9.0cm,angle=0]{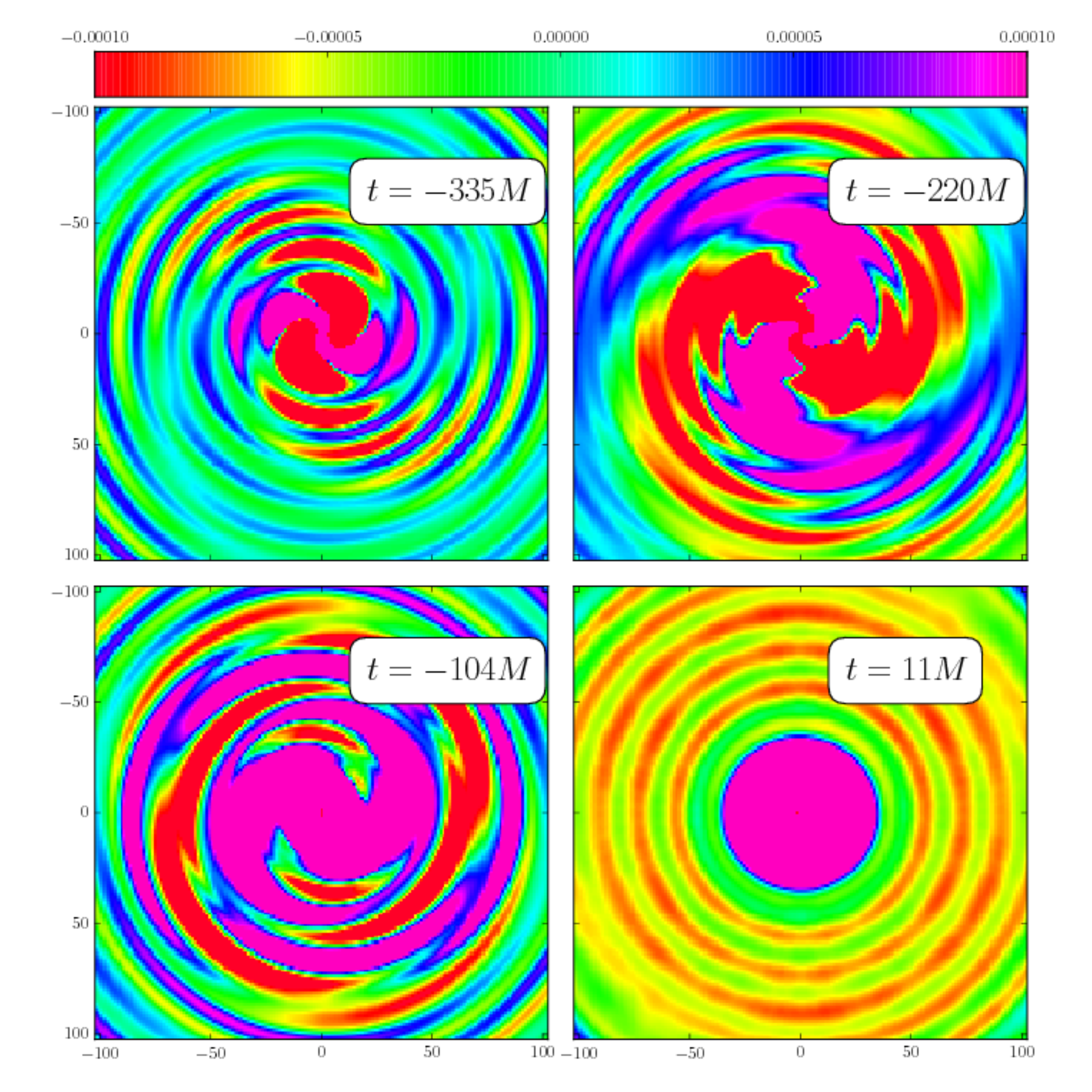}
\caption{ \textit{Difference in the conformal factor $\chi$ on the orbital plane between an inspiral with $A=0.5$, $R_0=0.3M$, and $M\Omega=0.5$ and the run without quantum fluctuations.}
The black hole binary rotates clockwise and has merged by the final frame shown.
The colormap is the same for all frames and has been chosen to emphasize the far zone. Note the change in character of the far-zone behavior after the merger, where the difference radiates symmetrically.
}
\label{fig:diff}
\end{figure}

In order to visualize the dynamical effects of these modifications, we
subtract the GR solution as a function of time from an evolution with fluctuations. 
Fig.~\ref{fig:diff} shows this difference for the conformal factor $\chi$ at four different times and illustrates the effect of the fluctuations.  Qualitatively, it indicates that the deviation from GR manifests dynamically as a coherent oscillating wave. During the inspiral, this oscillation appears roughly quadrupolar. In contrast,
after the merger, the amplitude decreases significantly and the oscillation is primarily monopolar.


\section{Observational Prospects}\label{observations}
%
We have shown qualitatively in Figs.~\ref{fig:orbital} and~\ref{fig:orbitalvaryomega} that quantum fluctuations modify the gravitational wave signal.  However, we need to quantify the extent to which these modified waveforms can be distinguished from the standard GR waveform by a gravitational wave detector such as aLIGO.

In order to assess the observability of the deviation, we compute the overlap between the perturbed waveforms and the standard GR waveform using methods from~\cite{Cutler:1994ys, Damour:1997ub, Fedrow:2017dpk}.  For two gravitational wave signals with strains $h_1(t)$ and $h_2(t)$, the overlap is given by the normalized inner product
\begin{eqnarray}
\label{overlapdef}
\mathcal O(1, 2) =  {\rm max}  \frac{\langle 1 | 2 \rangle}{\sqrt{\langle 1 | 1 \rangle \langle 2 | 2 \rangle}} 
\end{eqnarray}
maximized over time and phase shifts. In order to take into account the sensitivity of aLIGO over its bandwidth, 
we employ a frequency-domain, noise-weighted, inner product
\begin{eqnarray}
\label{innerprocudtdef}
\langle 1 | 2 \rangle = {\rm Re} \int_{-\infty}^\infty \frac{\tilde h_1(\omega) \tilde h_2^*(\omega)}{S(\omega)} \ ,
\end{eqnarray}
    where $\tilde h_1(\omega)$ and $\tilde h_2(\omega)$ are the Fourier transforms of $h_1(t)$ and $h_2(t)$.  The integrand is weighted by the noise profile of aLIGO $S(\omega)$, available at~\cite{LIGOdata}.\footnote{The LIGO noise profile ranges from 9 Hz to 8192 Hz, so in practice the integral in \eqref{innerprocudtdef} is cut off at those limits.}

Because  gravitational wave signals scale with the total mass $M$ of the binary pair, we must specify $M$ in order to determine the overlap across the sensitivity band of aLIGO.  Table \ref{tab:overlaps} presents the overlap $\mathcal O$ for all of the choices of fluctuation amplitude $A$ and frequency $M\Omega$ shown in Figs.~\ref{fig:orbital} and~\ref{fig:orbitalvaryomega}, each for several values of $M$.

\begin{table}[htbp]
   \centering
   \begin{tabular}{ll | ccc} 
                 &           &              &     Overlap $\mathcal O$ for &       \\  
       $A$   &  M$\Omega$ &  $M = 6 M_\odot$ &  $M = 20M_\odot$  & $M=60M_\odot$ \\  
      \hline
      0.05   &  0.1    &   0.997   &   0.996   &   0.996 \\
      0.1     &  0.1    &   0.999   &   0.999    &   0.999 \\
      0.5     &  0.1    &   0.694   &   0.764    &   0.827 \\ 
      0.5     &  0.25  &   0.757   &   0.802    &   0.840 \\
      0.5     &  0.5    &   0.492   &   0.599    &   0.692 \\
      0.5     &  0.75  &   0.517   &   0.616    &   0.702 \\
   \end{tabular}
   \caption{The overlap $\mathcal O$ between the GR waveform and the perturbed waveforms with amplitudes $A$ and frequencies $M \Omega$, computed using Eq.~\eqref{overlapdef}. } 
   \label{tab:overlaps}
\end{table}

As would be expected, fluctuations with larger amplitude have smaller overlap with the GR waveform.  Higher frequency fluctuations (at a fixed $M$) also leads to smaller overlap.

Varying the total mass has a notable effect on the overlap. aLIGO is most sensitive in a frequency band around 245 Hz, and for $M = 60M_\odot$, very roughly the total mass of GW150914~($65M_\odot$) and GW170104~($50M_\odot$),
the power spectrum of the signal is peaked at frequencies around 50~Hz.   Decreasing $M$ shifts the gravitational wave signal toward higher frequency and 
better sensitivity.  Consequently, overlaps are generally lower for smaller masses.

With these overlaps computed, we need to establish the threshold at which aLIGO is able to distinguish the two waveforms.  Ref.~\cite{Fedrow:2017dpk} argues that a mismatch $\mathcal M = 1-\mathcal O$ exceeding the inverse square of the signal-to-noise ratio will be observable. The signal to noise ratio for GW150914 was 24, implying that waveforms with overlap less than $0.998$ would be distinguishable. Here, we will adopt a more conservative standard of $\mathcal O \lesssim 0.97$ for distinguishability, a typical minimum overlap for neighboring points in template banks~\cite{Brown:2009ng,Chatziioannou:2013dza,Capano:2016dsf}, for which a signal to noise ratio of only 5.8 would be required.

Using this standard, Table~\ref{tab:overlaps} indicates that nominal amplitudes and frequencies
suggested by Giddings could be constrained by observations.
Several binary mergers have so far been detected with no observable deviation from the standard GR waveform. This implies, in principle, these aLIGO observations could set limits on the parameter space of the nonviolent nonlocality model. However, establishing such a bound will require addressing several challenges, which we outline below.

\section{Discussion}\label{discussion}
%
Our results indicate that the near-horizon fluctuations predicted by the strong 
nonviolent nonlocality proposal may significantly affect the orbital dynamics of inspiraling black hole binaries. Unlike the comparison between the geometry of a neutron star and that of a black hole of identical mass, these modifications to GR are radiative and impact the gravitational wave signal more when the black holes are well separated than just before they merge.  And, we have furthermore argued that aLIGO can, in principle, observe these effects.  We should note, however, a number of important caveats.

It is possible that the modifications to the gravitational wave signal resulting from the quantum fluctuations could alternatively be produced by changing other parameters, such as the black hole masses, spins, or separation.  In this case, even if the fluctuations did exist, the observed signal would be consistent with unmodified GR describing the inspiral of some alternate binary system. Conclusively ruling out such possible degeneracies would require a greatly extended overlap analysis across a range of parameters to determine the true distinguishability of the nonviolent nonlocality waveforms.  Such extensive numerical computations are beyond the scope of this work; our goal is a proof of principle that quantum fluctuations yield observable gravitational wave signals.

The nonviolent nonlocality model proposes that the black hole solution is modified by quantum metric fluctuations but does not  provide a consistent dynamical mechanism to produce them, such as a modification to Einstein's equation.  As a consequence, the procedure by which the quantum fluctuations are numerically implemented is necessarily ad hoc, in that no continuum limit is being solved.  As a result, alternative implementations are possible which might predict different or more subtle effects. For example, one could perturb the GR spacetime directly instead of the time derivatives.

Another alternative implementation could include an ensemble of fluctuations with different frequencies, 
which might yield more incoherent quantum effects.  In particular,  such a scheme might avoid
any peculiarities of a coherent, single-frequency fluctuation interacting with the orbiting system. We have implemented such an ensemble with random frequencies
chosen from a Gaussian distribution about a chosen central value and amplitudes
chosen such that the ``energies'' $A_i^2 \Omega_i^2$ of the fluctuations are also distributed in a Gaussian manner
about a central energy. Furthermore, the amplitudes are rescaled so that
the sum of the fluctuation energies matches the energy of a single mode. One
might guess that these modes might interfere in such a way that their effects would tend to cancel out. However, early results show a significant effect, so much so that the binary components are no longer bound and immediately fly apart. Similar behavior was observed for amplitudes larger than those presented here.

Due to the lack of a continuum limit, numerical effects are not well understood.  Although the behavior studied here does not appear sensitive to the resolution, moving away from the rigorous standard of
convergence and consistency leaves room for uncertainty.

Despite these caveats, we argue that the method described here, even if it falls short
of solving a fully nonlinear, consistent set of differential equations, represents a significant step toward a quantitative, observational test of the strong 
nonviolent nonlocality model. 
In this regard, our results suggest that such near-horizon fluctuations could be constrained
by gravitational wave observations of binary black hole mergers.

%
%
\vspace{0.5cm}

\begin{acknowledgments}
We thank Peter Shawhan for both helpful discussions and comments on
the manuscript. 
It is also a pleasure to thank Steve Giddings, Luis Lehner, Vassilios Mewes, Bob Wald, and Nico Yunes for helpfulcomments and discussions.
This work was supported by the NSF under grants PHY-1308621 \& PHY-1607291~(LIU) and
NASA's ATP program through grant NNX13AH01G.
Computations were
performed at XSEDE.
\end{acknowledgments}

\bibliographystyle{utphys}
\bibliography{paper}

\providecommand{\href}[2]{#2}\begingroup\raggedright\begin{thebibliography}{10}

\bibitem{Abbott:2016blz}
{\bfseries Virgo, LIGO Scientific} Collaboration, B.~â. Abbott {\em et~al.},
  ``{Observation of Gravitational Waves from a Binary Black Hole Merger},''
  \href{http://dx.doi.org/10.1103/PhysRevLett.116.061102}{{\em Phys. Rev.
  Lett.} {\bfseries 116} no.~6, (2016) 061102},
\href{http://arxiv.org/abs/1602.03837}{{\ttfamily arXiv:1602.03837 [gr-qc]}}.

\bibitem{Abbott:2016nmj}
{\bfseries Virgo, LIGO Scientific} Collaboration, B.~.~I. Abbott {\em et~al.},
  ``{GW151226: Observation of Gravitational Waves from a 22-Solar-Mass Binary
  Black Hole Coalescence},''
  \href{http://dx.doi.org/10.1103/PhysRevLett.116.241103}{{\em Phys. Rev.
  Lett.} {\bfseries 116} no.~24, (2016) 241103},
\href{http://arxiv.org/abs/1606.04855}{{\ttfamily arXiv:1606.04855 [gr-qc]}}.

\bibitem{Abbott:2017vtc}
{\bfseries Virgo} Collaboration, B.~P. Abbott {\em et~al.}, ``{GW170104:
  Observation of a 50-Solar-Mass Binary Black Hole Coalescence at Redshift
  0.2},'' \href{http://dx.doi.org/10.1103/PhysRevLett.118.221101}{{\em Phys.
  Rev. Lett.} {\bfseries 118} no.~22, (2017) 221101},
\href{http://arxiv.org/abs/1706.01812}{{\ttfamily arXiv:1706.01812 [gr-qc]}}.

\bibitem{TheLIGOScientific:2016src}
{\bfseries Virgo, LIGO Scientific} Collaboration, B.~P. Abbott {\em et~al.},
  ``{Tests of general relativity with GW150914},''
\href{http://arxiv.org/abs/1602.03841}{{\ttfamily arXiv:1602.03841 [gr-qc]}}.

\bibitem{Unruh:2017uaw}
W.~G. Unruh and R.~M. Wald, ``{Information Loss},''
\href{http://arxiv.org/abs/1703.02140}{{\ttfamily arXiv:1703.02140 [hep-th]}}.

\bibitem{Marolf:2017jkr}
D.~Marolf, ``{The Black Hole information problem: past, present, and future},''
\href{http://arxiv.org/abs/1703.02143}{{\ttfamily arXiv:1703.02143 [gr-qc]}}.

\bibitem{Page:1993wv}
D.~N. Page, ``{Information in black hole radiation},''
  \href{http://dx.doi.org/10.1103/PhysRevLett.71.3743}{{\em Phys. Rev. Lett.}
  {\bfseries 71} (1993) 3743--3746},
\href{http://arxiv.org/abs/hep-th/9306083}{{\ttfamily arXiv:hep-th/9306083
  [hep-th]}}.

\bibitem{Almheiri:2012rt}
A.~Almheiri, D.~Marolf, J.~Polchinski, and J.~Sully, ``{Black Holes:
  Complementarity or Firewalls?},''
  \href{http://dx.doi.org/10.1007/JHEP02(2013)062}{{\em JHEP} {\bfseries 02}
  (2013) 062},
\href{http://arxiv.org/abs/1207.3123}{{\ttfamily arXiv:1207.3123 [hep-th]}}.

\bibitem{Mathur:2005zp}
S.~D. Mathur, ``{The Fuzzball proposal for black holes: An Elementary
  review},'' \href{http://dx.doi.org/10.1002/prop.200410203}{{\em Fortsch.
  Phys.} {\bfseries 53} (2005) 793--827},
\href{http://arxiv.org/abs/hep-th/0502050}{{\ttfamily arXiv:hep-th/0502050
  [hep-th]}}.

\bibitem{Giddings:2011ks}
S.~B. Giddings, ``{Models for unitary black hole disintegration},''
  \href{http://dx.doi.org/10.1103/PhysRevD.85.044038}{{\em Phys. Rev.}
  {\bfseries D85} (2012) 044038},
\href{http://arxiv.org/abs/1108.2015}{{\ttfamily arXiv:1108.2015 [hep-th]}}.

\bibitem{Giddings:2012gc}
S.~B. Giddings, ``{Nonviolent nonlocality},''
  \href{http://dx.doi.org/10.1103/PhysRevD.88.064023}{{\em Phys. Rev.}
  {\bfseries D88} (2013) 064023},
\href{http://arxiv.org/abs/1211.7070}{{\ttfamily arXiv:1211.7070 [hep-th]}}.

\bibitem{Giddings:2014nla}
S.~B. Giddings, ``{Modulated Hawking radiation and a nonviolent channel for
  information release},''
  \href{http://dx.doi.org/10.1016/j.physletb.2014.08.070}{{\em Phys. Lett.}
  {\bfseries B738} (2014) 92--96},
\href{http://arxiv.org/abs/1401.5804}{{\ttfamily arXiv:1401.5804 [hep-th]}}.

\bibitem{Giddings:2014ova}
S.~B. Giddings, ``{Possible observational windows for quantum effects from
  black holes},'' \href{http://dx.doi.org/10.1103/PhysRevD.90.124033}{{\em
  Phys. Rev.} {\bfseries D90} no.~12, (2014) 124033},
\href{http://arxiv.org/abs/1406.7001}{{\ttfamily arXiv:1406.7001 [hep-th]}}.

\bibitem{Giddings:2016tla}
S.~B. Giddings, ``{Gravitational wave tests of quantum modifications to black
  hole structure -- with post-GW150914 update},''
  \href{http://dx.doi.org/10.1088/0264-9381/33/23/235010}{{\em Class. Quant.
  Grav.} {\bfseries 33} no.~23, (2016) 235010},
\href{http://arxiv.org/abs/1602.03622}{{\ttfamily arXiv:1602.03622 [gr-qc]}}.

\bibitem{Giddings:2017mym}
S.~B. Giddings, ``{Nonviolent unitarization: basic postulates to soft quantum
  structure of black holes},''
\href{http://arxiv.org/abs/1701.08765}{{\ttfamily arXiv:1701.08765 [hep-th]}}.

\bibitem{Yosifov:2017pxj}
A.~Y. Yosifov and L.~G. Filipov, ``{Entropic Entanglement: Information Prison
  Break},''
\href{http://arxiv.org/abs/1707.01768}{{\ttfamily arXiv:1707.01768 [gr-qc]}}.

\bibitem{Giddings:2016btb}
S.~B. Giddings and D.~Psaltis, ``{Event Horizon Telescope Observations as
  Probes for Quantum Structure of Astrophysical Black Holes},''
\href{http://arxiv.org/abs/1606.07814}{{\ttfamily arXiv:1606.07814
  [astro-ph.HE]}}.

\bibitem{Estes:2016wgv}
J.~Estes, M.~Kavic, M.~Lippert, and J.~H. Simonetti, ``{Shining Light on
  Quantum Gravity with Pulsar-Black Hole Binaries},''
  \href{http://dx.doi.org/10.3847/1538-4357/aa610e}{{\em Astrophys. J.}
  {\bfseries 837} no.~1, (2017) 87},
\href{http://arxiv.org/abs/1607.00018}{{\ttfamily arXiv:1607.00018 [hep-th]}}.

\bibitem{emd}
E.~W. Hirschmann, L.~Lehner, S.~L. Liebling, and C.~Palenzuela, ``{Black Hole
  Dynamics in Einstein-Maxwell-Dilaton Theory},''
\href{http://arxiv.org/abs/1706.09875}{{\ttfamily arXiv:1706.09875 [gr-qc]}}.

\bibitem{Okounkova:2017yby}
M.~Okounkova, L.~C. Stein, M.~A. Scheel, and D.~A. Hemberger, ``{Numerical
  binary black hole mergers in dynamical Chern-Simons: I. Scalar field},''
\href{http://arxiv.org/abs/1705.07924}{{\ttfamily arXiv:1705.07924 [gr-qc]}}.

\bibitem{Barcelo:2014cla}
C.~Barcelo, R.~Carballo-Rubio, L.~J. Garay, and G.~Jannes, ``{The lifetime
  problem of evaporating black holes: mutiny or resignation},''
  \href{http://dx.doi.org/10.1088/0264-9381/32/3/035012}{{\em Class. Quant.
  Grav.} {\bfseries 32} no.~3, (2015) 035012},
\href{http://arxiv.org/abs/1409.1501}{{\ttfamily arXiv:1409.1501 [gr-qc]}}.

\bibitem{Cardoso:2016oxy}
V.~Cardoso, S.~Hopper, C.~F.~B. Macedo, C.~Palenzuela, and P.~Pani,
  ``{Gravitational-wave signatures of exotic compact objects and of quantum
  corrections at the horizon scale},''
  \href{http://dx.doi.org/10.1103/PhysRevD.94.084031}{{\em Phys. Rev.}
  {\bfseries D94} no.~8, (2016) 084031},
\href{http://arxiv.org/abs/1608.08637}{{\ttfamily arXiv:1608.08637 [gr-qc]}}.

\bibitem{Maggio:2017ivp}
E.~Maggio, P.~Pani, and V.~Ferrari, ``{Exotic Compact Objects and How to Quench
  their Ergoregion Instability},''
\href{http://arxiv.org/abs/1703.03696}{{\ttfamily arXiv:1703.03696 [gr-qc]}}.

\bibitem{2017arXiv170606155M}
Z.~{Mark}, A.~{Zimmerman}, S.~M. {Du}, and Y.~{Chen}, ``{A recipe for echoes
  from exotic compact objects},'' {\em ArXiv e-prints} (June, 2017) ,
  \href{http://arxiv.org/abs/1706.06155}{{\ttfamily arXiv:1706.06155 [gr-qc]}}.

\bibitem{Abedi:2016hgu}
J.~Abedi, H.~Dykaar, and N.~Afshordi, ``{Echoes from the Abyss: Evidence for
  Planck-scale structure at black hole horizons},''
\href{http://arxiv.org/abs/1612.00266}{{\ttfamily arXiv:1612.00266 [gr-qc]}}.

\bibitem{Ashton:2016xff}
G.~Ashton, O.~Birnholtz, M.~Cabero, C.~Capano, T.~Dent, B.~Krishnan, G.~D.
  Meadors, A.~B. Nielsen, A.~Nitz, and J.~Westerweck, ``{Comments on: "Echoes
  from the abyss: Evidence for Planck-scale structure at black hole
  horizons"},''
\href{http://arxiv.org/abs/1612.05625}{{\ttfamily arXiv:1612.05625 [gr-qc]}}.

\bibitem{Abedi:2017isz}
J.~Abedi, H.~Dykaar, and N.~Afshordi, ``{Echoes from the Abyss: The Holiday
  Edition!},''
\href{http://arxiv.org/abs/1701.03485}{{\ttfamily arXiv:1701.03485 [gr-qc]}}.

\bibitem{Barcelo:2017lnx}
C.~Barceló, R.~Carballo-Rubio, and L.~J. Garay, ``{Gravitational wave echoes
  from macroscopic quantum gravity effects},''
  \href{http://dx.doi.org/10.1007/JHEP05(2017)054}{{\em JHEP} {\bfseries 05}
  (2017) 054},
\href{http://arxiv.org/abs/1701.09156}{{\ttfamily arXiv:1701.09156 [gr-qc]}}.

\bibitem{Price:2017cjr}
R.~Price and G.~Khanna, ``{Gravitational wave sources: reflections and
  echoes},''
\href{http://arxiv.org/abs/1702.04833}{{\ttfamily arXiv:1702.04833 [gr-qc]}}.

\bibitem{Hawking:1987en}
T.~Damour, ``The problem of motion in newtonian and einsteinian gravity,'' in
  {\em Three Hundred Years of Gravitation}, S.~Hawking and W.~Israel, eds.,
  pp.~128--198.
\newblock Cambridge University Press, Cambridge; New York, 1987.

\bibitem{1995PhRvD..52.5428S}
M.~{Shibata} and T.~{Nakamura}, ``{Evolution of three-dimensional gravitational
  waves: Harmonic slicing case},''
  \href{http://dx.doi.org/10.1103/PhysRevD.52.5428}{{\em Phys. Rev. D}
  {\bfseries 52} (Nov., 1995) 5428--5444}.

\bibitem{1999PhRvD..59b4007B}
T.~W. {Baumgarte} and S.~L. {Shapiro}, ``{Numerical integration of Einstein's
  field equations},'' \href{http://dx.doi.org/10.1103/PhysRevD.59.024007}{{\em
  Phys. Rev. D} {\bfseries 59} no.~2, (Jan., 1999) 024007},
  \href{http://arxiv.org/abs/gr-qc/9810065}{{\ttfamily gr-qc/9810065}}.

\bibitem{Palenzuela:2009yr}
C.~Palenzuela, M.~Anderson, L.~Lehner, S.~L. Liebling, and D.~Neilsen,
  ``{Stirring, not shaking: binary black holes' effects on electromagnetic
  fields},'' \href{http://dx.doi.org/10.1103/PhysRevLett.103.081101}{{\em Phys.
  Rev. Lett.} {\bfseries 103} (2009) 081101},
\href{http://arxiv.org/abs/0905.1121}{{\ttfamily arXiv:0905.1121
  [astro-ph.HE]}}.

\bibitem{Neilsen:2010ax}
D.~Neilsen, L.~Lehner, C.~Palenzuela, E.~W. Hirschmann, S.~L. Liebling, {\em
  et~al.}, ``{Boosting jet power in black hole spacetimes},''
  \href{http://dx.doi.org/10.1073/pnas.1019618108}{{\em Proc.Nat.Acad.Sci.}
  {\bfseries 108} (2011) 12641--12646},
\href{http://arxiv.org/abs/1012.5661}{{\ttfamily arXiv:1012.5661
  [astro-ph.HE]}}.

\bibitem{Liebling:2016orx}
S.~L. Liebling and C.~Palenzuela, ``{Electromagnetic Luminosity of the
  Coalescence of Charged Black Hole Binaries},''
  \href{http://dx.doi.org/10.1103/PhysRevD.94.064046}{{\em Phys. Rev.}
  {\bfseries D94} no.~6, (2016) 064046},
\href{http://arxiv.org/abs/1607.02140}{{\ttfamily arXiv:1607.02140 [gr-qc]}}.

\bibitem{1987PThPS..90....1N}
T.~{Nakamura}, K.~{Oohara}, and Y.~{Kojima}, ``{General Relativistic Collapse
  to Black Holes and Gravitational Waves from Black Holes},''
  \href{http://dx.doi.org/10.1143/PTPS.90.1}{{\em Progress of Theoretical
  Physics Supplement} {\bfseries 90} (1987) 1--218}.

\bibitem{lousto}
M.~Campanelli, C.~O. Lousto, P.~Marronetti, and Y.~Zlochower, ``Accurate
  evolutions of orbiting black-hole binaries without excision,'' {\em Phys.
  Rev. Lett.} {\bfseries 96} (2006) 111101,
\href{http://arxiv.org/abs/gr-qc/0511048}{{\ttfamily gr-qc/0511048}}.

\bibitem{Hannam:2009ib}
M.~Hannam, S.~Husa, and N.~O. Murchadha, ``{Bowen-York trumpet data and
  black-hole simulations},''
  \href{http://dx.doi.org/10.1103/PhysRevD.80.124007}{{\em Phys. Rev.}
  {\bfseries D80} (2009) 124007},
\href{http://arxiv.org/abs/0908.1063}{{\ttfamily arXiv:0908.1063 [gr-qc]}}.

\bibitem{Cutler:1994ys}
C.~Cutler and E.~E. Flanagan, ``{Gravitational waves from merging compact
  binaries: How accurately can one extract the binary's parameters from the
  inspiral wave form?},''
  \href{http://dx.doi.org/10.1103/PhysRevD.49.2658}{{\em Phys. Rev.} {\bfseries
  D49} (1994) 2658--2697},
\href{http://arxiv.org/abs/gr-qc/9402014}{{\ttfamily arXiv:gr-qc/9402014
  [gr-qc]}}.

\bibitem{Damour:1997ub}
T.~Damour, B.~R. Iyer, and B.~S. Sathyaprakash, ``{Improved filters for
  gravitational waves from inspiralling compact binaries},''
  \href{http://dx.doi.org/10.1103/PhysRevD.57.885}{{\em Phys. Rev.} {\bfseries
  D57} (1998) 885--907},
\href{http://arxiv.org/abs/gr-qc/9708034}{{\ttfamily arXiv:gr-qc/9708034
  [gr-qc]}}.

\bibitem{Fedrow:2017dpk}
J.~M. Fedrow, C.~D. Ott, U.~Sperhake, J.~Blackman, R.~Haas, C.~Reisswig, and
  A.~De~Felice, ``{Gravitational Waves from Binary Black Hole Mergers Inside of
  Stars},''
\href{http://arxiv.org/abs/1704.07383}{{\ttfamily arXiv:1704.07383
  [astro-ph.HE]}}.

\bibitem{LIGOdata}
 \url{https://dcc.ligo.org/LIGO-T0900288/public}. Specifically the ``Zero Det.
  High Power'' curve.

\bibitem{Brown:2009ng}
D.~A. Brown and P.~J. Zimmerman, ``{The Effect of Eccentricity on Searches for
  Gravitational-Waves from Coalescing Compact Binaries in Ground-based
  Detectors},'' \href{http://dx.doi.org/10.1103/PhysRevD.81.024007}{{\em Phys.
  Rev.} {\bfseries D81} (2010) 024007},
\href{http://arxiv.org/abs/0909.0066}{{\ttfamily arXiv:0909.0066 [gr-qc]}}.

\bibitem{Chatziioannou:2013dza}
K.~Chatziioannou, A.~Klein, N.~Yunes, and N.~Cornish, ``{Gravitational
  Waveforms for Precessing, Quasicircular Compact Binaries with Multiple Scale
  Analysis: Small Spin Expansion},''
  \href{http://dx.doi.org/10.1103/PhysRevD.88.063011}{{\em Phys. Rev.}
  {\bfseries D88} no.~6, (2013) 063011},
\href{http://arxiv.org/abs/1307.4418}{{\ttfamily arXiv:1307.4418 [gr-qc]}}.

\bibitem{Capano:2016dsf}
C.~Capano, I.~Harry, S.~Privitera, and A.~Buonanno, ``{Implementing a search
  for gravitational waves from binary black holes with nonprecessing spin},''
  \href{http://dx.doi.org/10.1103/PhysRevD.93.124007}{{\em Phys. Rev.}
  {\bfseries D93} no.~12, (2016) 124007},
\href{http://arxiv.org/abs/1602.03509}{{\ttfamily arXiv:1602.03509 [gr-qc]}}.

\end{thebibliography}\endgroup

\end{document}